\documentstyle [12pt,twoside,psfig]{article}

\newcommand{\bq}{\begin{equation}}
\newcommand{\ba}{\begin{eqnarray}}
\newcommand{\eq}{\end{equation}}
\newcommand{\ea}{\end{eqnarray}}

\def\a{\alpha}
\def\b{\beta}

\def\d{\delta}

\def\f{\phi}
\def\g{\raisebox{.4ex}{$\gamma$}}

\def\l{\lambda}

\def\p{\pi}

\def\t{\tau}

\def\F{\Phi}

\def\J{\Psi}

\def\bo{{\raise.15ex\hbox{\large$\Box$}}}
\def\bob{{\lower.2ex\hbox{\large$\Box$}}}
\def\pa{\partial}

\def\TH{{\raise.2ex\hbox{$\displaystyle \bigodot$}\mskip-4.7mu \llap H \;}}
 
\def\bra#1{\left\langle #1\right|}
\def\ket#1{\left| #1\right\rangle}

\def\VEV#1{\left\langle #1\right\rangle}
 
\def\abs#1{\left| #1\right|}
 
\catcode`@=11
\def\underline#1{\relax\ifmmode\@@underline#1\else
        $\@@underline{\hbox{#1}}$\relax\fi}
\catcode`@=12
 
\begin{document}

\hfill LA-UR-94-4318
\centerline{\large{\bf Statistical Mechanics of Nonlinear Coherent
Structures:}}
\centerline{\large{\bf Kinks in the $\Phi^6$
Model \footnotemark}}
\footnotetext{To appear in Proceedings of NEEDS'94, Los Alamos,
September (1994)}  

\vspace{2cm}

\centerline{\bf{Salman Habib and Avadh Saxena}}

\vspace{1cm}

\centerline{\em{Theoretical Division}}
\centerline{\em{Los Alamos National Laboratory}}
\centerline{\em{Los Alamos, NM 87545}}  

\vspace{2cm}

\centerline{\bf Abstract}

We study the thermodynamics of kinks in the $\Phi^6$ model using a
Langevin code implemented on a massively parallel computer. This code
can be used to study first order dynamical phase transitions which
exhibit multiple length and time scales. The classical statistical
mechanics of a $1+1$-dimensional field theory reduces to a
time-independent quantum problem in one dimension via the transfer
integral method. Exact solutions of the Schr\"odinger equation exist
for the $\Phi^6$ potential (unlike the case for $\Phi^4$) and can be
used to check results from the simulations. The $\Phi^6$ model is also
much richer than the $\Phi^4$ model in terms of the variety of
coherent structures and possible phases accompanying a phase
transition. Specifically, we have calculated (in a one dimensional
model) such quantities as the probability density function (PDF) and
field-field correlation functions. These quantities help us understand
the contribution to the specific heat from coherent structures such as
domain walls (kinks) and other transformation structures as opposed to
the contribution from lattice vibrations.  We have calibrated our results
against known exact solutions for limiting cases with very high
accuracy. Having understood this problem, we are now extending our
Langevin code to higher dimensions.
 
\newpage

\section{Introduction} 

First order phase transitions are ubiquitous in nature, ranging from
melting to structural transformations in crystals. Due to the
discontinuity in the order parameter at a first order transition, a
simple symmetric double well potential is incapable of describing such
a transition and an asymmetric double well or a symmetric triple well
potential (as, {\em e.g.}, a $\Phi^6$ model) is needed. A symmetric
double well potential ($\Phi^4$ model) is usually employed to describe
continuous (second order) transitions. The $\Phi^4$ model and its
attendant kink structure has been extensively studied in the
literature using techniques such as the path integral formalism
\cite{KS}, Langevin dynamics \cite{AHK}, {\em etc}. The $\Phi^6$ model,
appropriate for first order phase transitions, is much richer in terms
of its kink structure (``textures'' in the materials context)
\cite{BK1}\cite{FF}\cite{BK2}\cite{VGM}. However, unlike the $\Phi^4$ model,
many aspects of the $\Phi^6$ model, specifically the variation of the
probability density function (PDF) of the order parameter \cite{ADB},
correlation functions and structure factor within the context of
Langevin dynamics, have not been studied (except in a specific case
\cite{MG}). 

Here we adopt a novel viewpoint to study the thermodynamics of a
system based almost entirely on its PDF, {\em i.e.}, if the PDF is
known then the ground state eigenfunction and eigenvalue are known,
and therefore thermodynamic quantities, {\em e.g.}, the specific heat,
can be determined. (Correlation functions can also be determined but
not directly from the PDF.) It is clear that the PDF must be
determined to a very high accuracy. Our Langevin code, implemented on
a massively parallel computer, computes the PDF to the required
precision (in an appropriate temperature range). This can be easily
checked for the $\Phi^6$ model because certain exact solutions of the
Schr\"odinger equation with a $\Phi^6$ potential exist
\cite{BK1}\cite{SBD}\cite{GPF}. Indeed, we find spectacular agreement
between the computed and exact results for the $\Phi^6$ model. A
direct calibration of the Langevin code in this manner against the
$\Phi^4$ model is not possible because the Schr\"odinger equation with
a $\Phi^4$ potential has no exact solution. In addition, the field
configuration for a kink can be calibrated against the exact solution
for the $\Phi^6$ kink. In this case too we find very good agreement
between the Langevin code and the exact solution.

\section{The $\Phi^6$ Model} 

The Landau free energy density for the $\Phi^6$ model in one dimension
is given by   
\bq
{\bar{F}}_L({\bar{\F}})={1\over 2} A{\bar{\F}}^2+{1\over
4}B{\bar{\F}}^4+{1\over 6}{\bar{\F}^6}~,                   \label{1} 
\eq
where to describe a first order (or discontinuous) phase transition it
is necessary that $B<0$ and $C>0$. In many physical cases ({\em e.g.},
soft phonon mode driven structural transformations) $A$ is chosen to
be temperature dependent \cite{FF}\cite{BK2}. This model also exhibits
a second order (or continuous) transition for $B>0$, $C>0$ at $A = 0$.
To include domain walls (of nonvanishing width and energy) between
various phases the Landau free energy is supplemented by adding the
square of the field gradient (the Ginzburg term)  
\bq
\bar{F}_{GL}(\F,{\bar{\F}}^{\prime})=F_L(\bar{\F})+{1\over 2} d
{\bar{\F}}^{{\prime}2}~,                                       \label{2} 
\eq
where ${\bar{\F}}^{\prime}=\pa\bar{\F}/\pa \bar{x}$ and $d>0$. The
coefficient $d$ also pertains to a physical quantity such as the soft
shear modulus of a crystal \cite{FF}\cite{BK2}. For convenience we use
dimensionless variables \cite{BK2} for the Ginzburg-Landau energy
$F_{GL}$, the field $\Phi$, and the spatial variable $x$ according to    
\ba
\bar{F}_{GL}&=&\l F_{GL}~,~~~~~\l=-{9 B^3\over 16 C^2}~,   \label{3}\\
\bar{\F}&=&\g\F~,~~~~~\g=\left(-{3 B\over 2C}\right)^{1/2}~,
                                                           \label{4}\\
\bar{x}&=&\nu x~,~~~~~\nu=\left({4 dC\over 3B^2}\right)^{1/2}~.
                                                           \label{5}
\ea
Thus, in dimensionless form we get 
\bq
F_{GL}={1\over 4}\t\F^2-\F^4+\F^6+{\F}^{{\prime}2}~,           \label{6}
\eq
where the prime denotes differentiation with respect to $x$, and 
\bq
\t={16AC\over 3B^2}~,                                      \label{7}
\eq
can be interpreted as a dimensionless temperature. The $\Phi^6$
potential (\ref{1}) has three minima that occur at 
\ba
\bar{\F}&=&0~,~~~~~\bar{\F}_{min}=\pm\bar{\F}_0(1+z)^{1/2}~,\label{8}\\
\bar{\F}_0&=&{1\over\sqrt{3}}\g~,~~~~~z=\left(1-{3\over
4}\t\right)^{1/2}~,                                         \label{9}
\ea 
and two maxima that occur at 
\bq
\bar{\F}_{max}=\pm\bar{\F}_0(1-z)^{1/2}~.                   \label{10}
\eq
The value of the potential (Landau free energy) at the extrema is
given by 
\bq 
\bar{F}_{\pm}=\bar{F}_0(1\mp z)^2(1\pm 2z)~,                \label{11}
\eq
where $\bar{F}_+$ represents a free energy maximum, $\bar{F}_-$, a free
energy minimum, and $\bar{F}_0=(2/27)\l$. For the dimensionless free
energy F$_{GL}$ and field $\Phi$ one just sets $\g = \l = \nu = 1$ in
Eqs. (\ref{8}-\ref{11}).

Physically the central minimum represents the high temperature
(parent) phase while the two side minima represent the two variants of
a fully developed low temperature (product) phase. Similarly, the two
maxima refer to the two variants of a partially developed product
phase. For $\tau < 0$ there are only two minima and the free energy
behaves as an effective $\Phi^4$ model. For $0 < \tau < 4/3$ there are
three minima. At $\tau = 1$ there are three degenerate minima which
correspond to the first order transition point, {\em i.e.}, the parent
and product phases coexist in equilibrium.  For $\tau > 4/3$ there is
a single well corresponding to the fact that only the parent phase is
stable. At $\tau = 4/3$ there is only one minimum at $\Phi = 0$ and
two points of inflection at $\pm (-B/2C)^{1/2}$.

\section{Exact Kink and Domain Wall Solutions}
 
The equilibrium field configuration is determined by minimizing the
total free energy. The Euler-Lagrange equations after two integrations
lead to 
\bq  
x(\F)={1\over 2}\int{d\f\over\sqrt{\f\left(\f^3-\f^2+{\t\over
4}\f-F_0\right)}}~,                                         \label{12}
\eq
with $\phi$ = $\Phi^2$ and the boundary conditions $F_0 = \lim_{x\rightarrow 
\pm\infty} F_L(x),~(F_L(x) > F_0)$, $\lim_{x\rightarrow \pm\infty}
{\Phi}^{\prime} = 0$. The following four kink and domain wall
solutions exist.\\
\\
\noindent{\bf 1.} For $\tau < 1$ and $F_0 < 0$ a kink solution between
the two product variants is given by 
\bq
\F(x)={\F_{min}\a\sinh{\b x}\over\left(\b^2+\a^2\sinh^2{\b
x}\right)^{1/2}}~,                                          \label{13}
\eq
where
\bq
\a=\F_{min}\abs{2\F_{min}^2-1}^{1/2}~,~~~~~\b=\F_{min}\left
(3\F_{min}^2-1\right)^{1/2}~.                               \label{14}
\eq
This kink (Fig. 1) corresponds to the $\Phi^4$ kink for $\tau < 0$.\\

\begin{figure}
\centerline{\psfig{figure=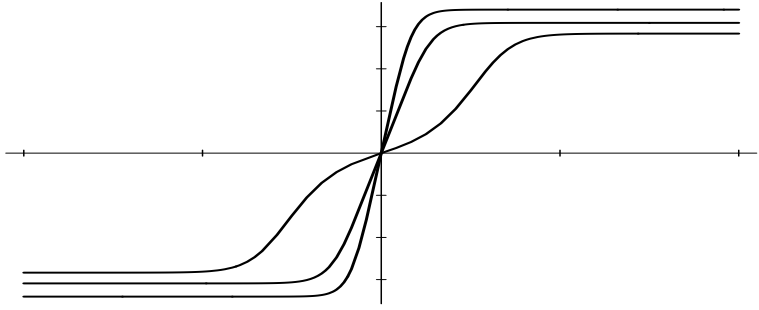,height=9cm,width=10cm}}
\caption[Figure 1]{\small{Kink solutions of type {\bf 1} for
$\tau=.99,~.5,~-.5$ (asymptotically, the bottom, middle, and top
curves). Note the ``sticky'' behavior of the solution near $\Phi=0$
for $\tau=.99$.}} 
\end{figure}  

\noindent{\bf 2.} For $1 < \tau < 4/3$ and $F_0 > 0$  a pulse soliton
between the parent and either product variant (with the parent phase
in the middle) is given by  
\bq
\F(x)={\F_{min}\a\over\left(\F_{min}^4-\b^2\tanh^2{\b x}\right)^{1/2}}~. 
                                                             \label{15}
\eq
This pulse solution is depicted in Fig. 2. The total free energy for
solutions {\bf 1} and {\bf 2} is given by 
\bq
F_{tot}={1\over
2}\left(\b-(\t-1)\ln\left({\b+\F_{min}^2\over\a}\right)\right)~.
                                                             \label{16} 
\eq

\begin{figure}
\centerline{\psfig{figure=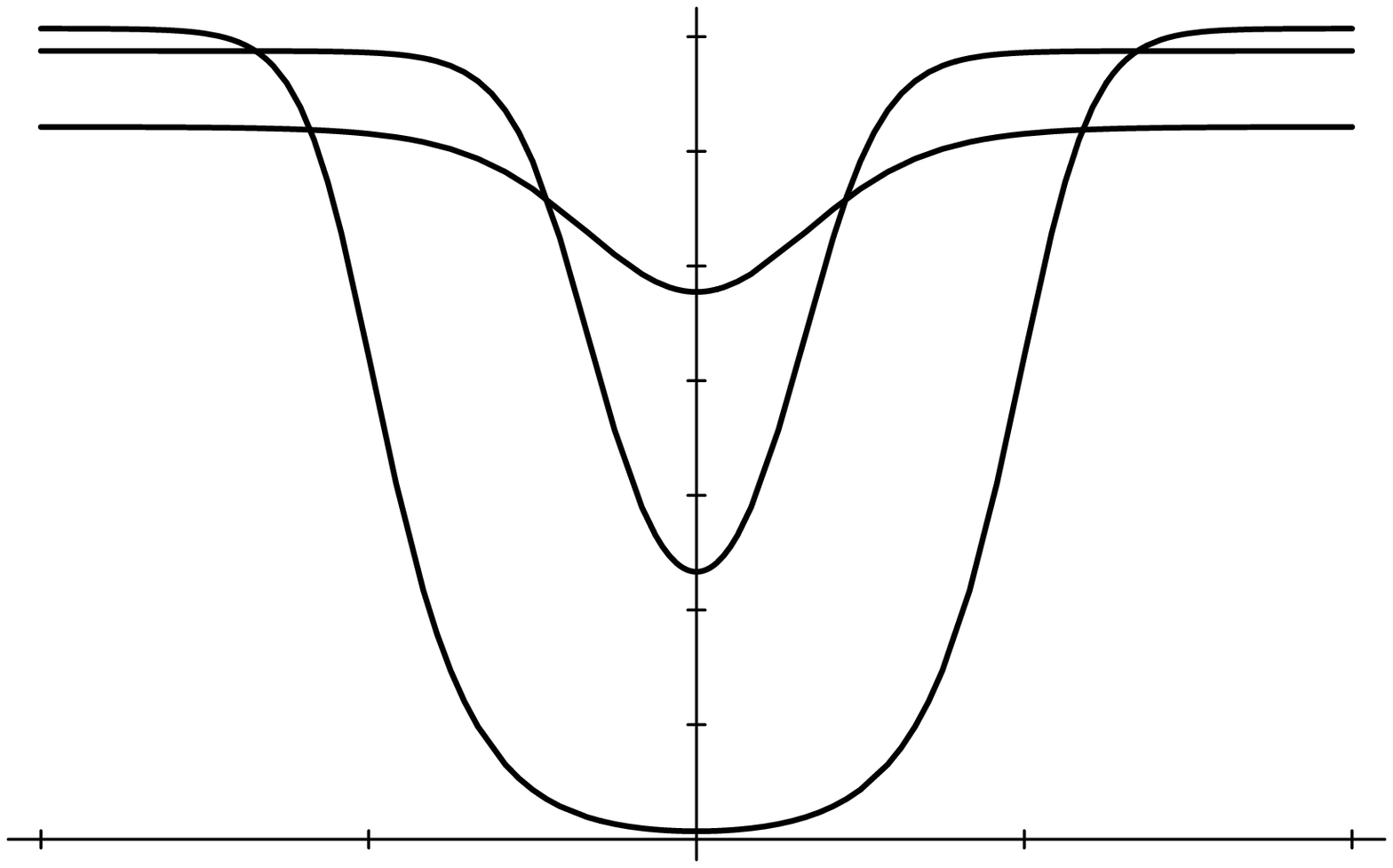,height=9cm,width=10cm}}
\caption[Figure 2]{\small{Pulse solutions of type {\bf 2} for
$\tau=1.0001,~1.1,~1.3$. (asymptotically, the bottom, middle, and top
curves)}} 
\end{figure} 
 
\noindent{\bf 3.} For $0 < \tau < 1$ and $F_0 = 0$ a pulse soliton
between either product variant and the parent phase (with either
product variant in the middle) is given by 
\bq
\F(x)={\F_2\over\left(1+\left(1-{\F_2^2\over\F_3^2}\right)
\sinh^2{\sqrt{\t}x/2}\right)^{1/2}}~,                   \label{17} 
\eq
where $\F_{2,3}^2=(1/2)(1\mp(1-\t)^{1/2})$. This solution is shown in
Fig. 3. The total free energy for this pulse solution is given by 
\bq
F_{tot}={1\over 4}\left(1-(1-\t)\ln\left({\sqrt{2}(1+\sqrt{\t})
\over\sqrt{1-\t}}\right)\right)~.                            \label{18} 
\eq

\begin{figure}
\centerline{\psfig{figure=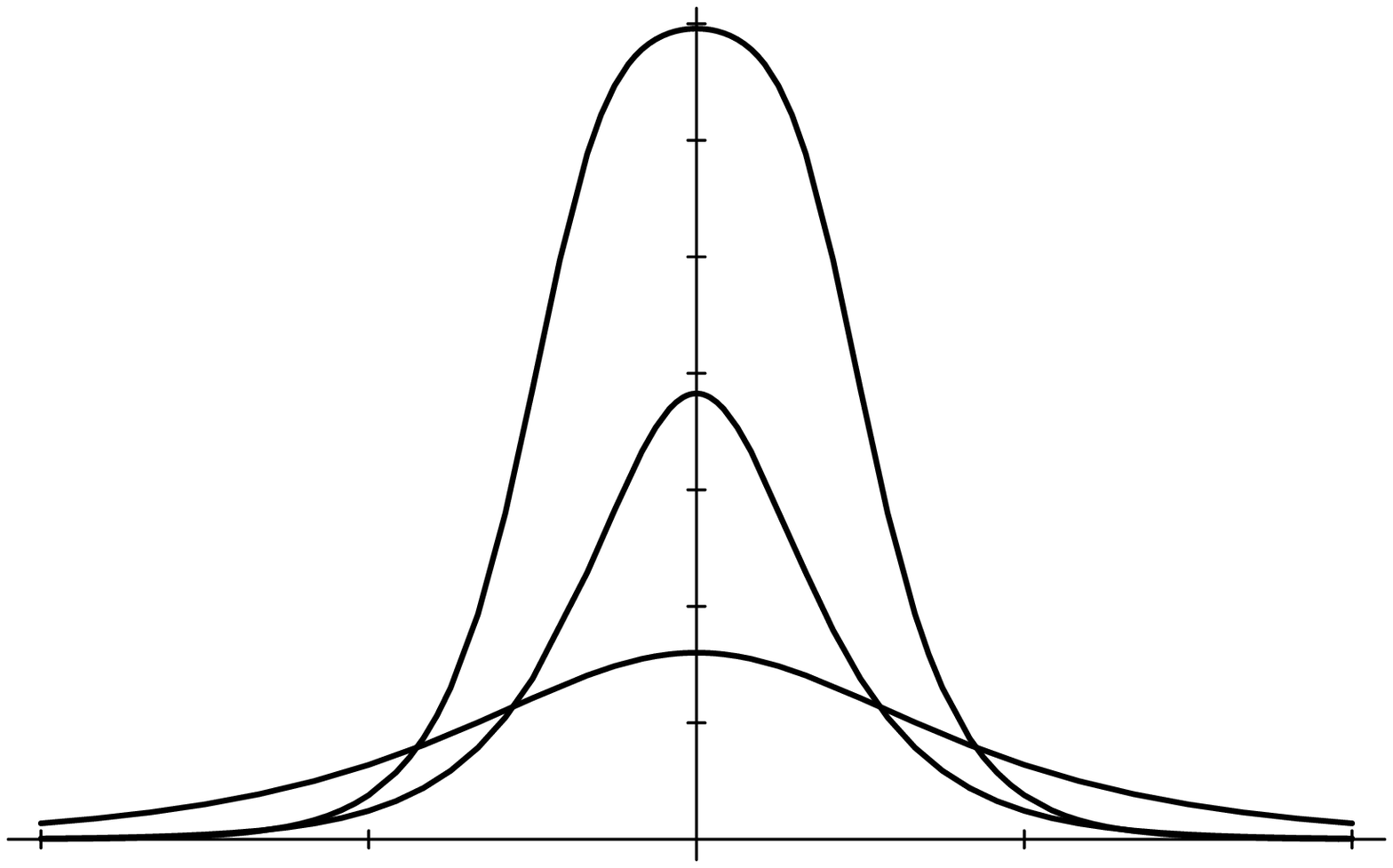,height=9cm,width=10cm}}
\caption[Figure 3]{\small{Pulse solutions of type {\bf 3} for
$\tau=.1,~.5,~.999$ (the top, middle, and bottom peaks
respectively).}} 
\end{figure} 
 
\noindent{\bf 4.} For $\tau = 1$ and $F_0 = 0$ a ``half'' kink between
either product variant and the parent phase (at the transition point)
is given by (Fig. 4)
\bq
\F(x)={1\over\sqrt{2}}\left(1+\hbox{e}^{-x}\right)^{-1/2}~,  \label{19}
\eq
with total energy $F_{tot} = 1/8$.  This is the limiting case of {\bf
3} (matching with half a pulse of type {\bf 3}). Note that the above
four solutions are known in the literature in a different form
(and materials context) \cite{BK1}\cite{FF}\cite{VGM}\cite{FF2}\cite{JL}. 

\begin{figure}
\centerline{\psfig{figure=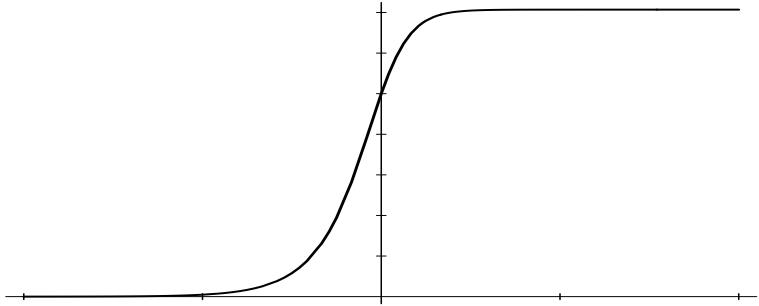,height=9cm,width=10cm}}
\caption[Figure 4]{\small{Half kink solution of type {\bf 4}.}}
\end{figure}     

\section{Langevin Dynamics on a Parallel Machine}

Solution of field theoretic Langevin equations is particularly
convenient on a massively parallel computer because little
inter-processor communication is involved and because the large memory
available enables the use of large lattices. For the results reported
here, lattice sizes ranged from several thousands to hundreds of
thousands of lattice points. The system size was kept much larger than
the field correlation length (at least a factor of ten, but much
larger typically). Periodic boundary conditions were used for
convenience.

The Langevin equation for the $\F^6$ model is 
\bq
\pa_{tt}^2\F=\pa_{xx}^2\F-\eta\pa_t\F+\F(1-\F^2)+\F^5+{\hat F}(x,t)~,
                                                            \label{le}
\eq
where the viscosity $\eta$ and the Gaussian white noise ${\hat F}$ are
related by the fluctuation-dissipation theorem:
\bq
\VEV{{\hat F}(x,t){\hat
F}(x^{\prime},t^{\prime})}=2\eta\beta^{-1}\d(x-x^{\prime}) 
\d(t-t^{\prime})~.                                          \label{fdt}
\eq
The lattice versions of the above continuous equations were then
solved using standard techniques \cite{LE}. Random initial conditions
were driven to equilibrium and the results sampled in time thereafter
to yield time averaged PDF's, {\em etc.} The use of the Langevin
technique for obtaining thermodynamic quantities is straightforward
and remarkably accurate. Moreover, structures such as the various kink
solutions can be clearly identified and information about real time
dynamical quantities such as the temporal correlation functions is
also available. 

\section{Results} 
\subsection{Probability Distribution Function}

The exact solution for the PDF is given by (the ground state wave
function \cite{SBD}\cite{GPF} squared for the quantum mechanical
problem is the PDF) 
\bq
\J_0^2=N\exp\left\{-\b\left[{1\over 2}\left({dC\over
3}\right)^{1/2}\F^4 + {B \over 4}\left({3d\over
C}\right)^{1/2}\F^2\right]\right\} 
\eq
where $N$ is a normalization constant. The corresponding ground state
energy \cite{SBD}\cite{GPF} is 
\bq
E_0={B\over 8\b}\left(3\over dC\right)^{1/2}.
\eq
In Fig. 5 we compare the probability density (PDF) function computed
from a Langevin simulation (diamonds) with the exact solution
(dashed line). The choice of parameters is $A=B=d=1$, $C=0.17$, and
$1/\b=0.144146$. As is apparent, the agreement is excellent, and is
indicative of the very high accuracy of the Langevin simulations. The
results of our computations are, for practical purposes, numerically
exact. Note that such a comparison is not directly possible in the
case of the $\Phi^4$ model \cite{AHK} since there are no exact
solutions in that case. 

\begin{figure}
\centerline{\psfig{figure=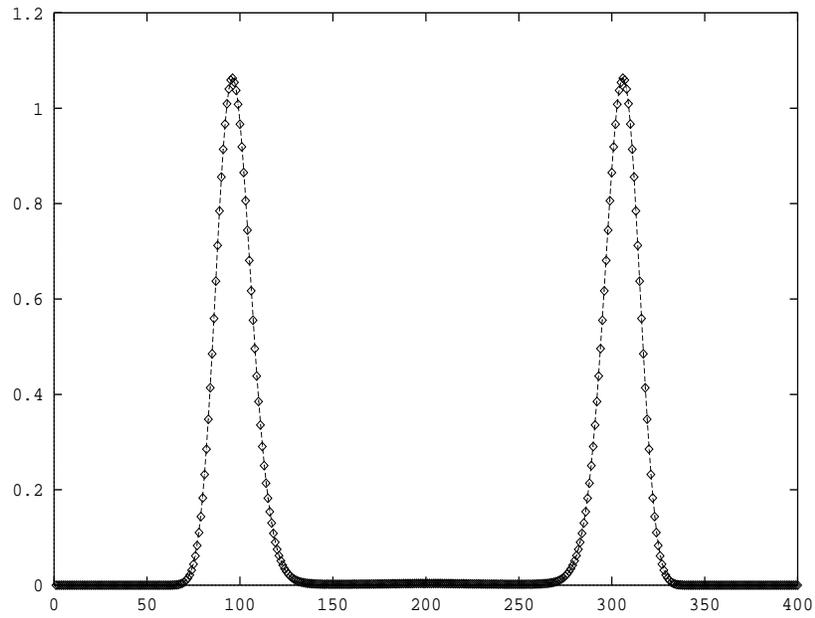,height=8cm,width=10cm}}
\caption[Figure 5]{\small{Comparison of the exact (dashed line) and
numerical (diamonds) PDF. The agreement is excellent (parameter values
are given in the text).}}
\end{figure}     

We have studied the variation of the PDF with temperature for several
temperatures ranging from well below the ``transition'' temperature to
well above it. In general, the PDF exhibits a three peak structure.
However, at a certain temperature the three peaks have the same
height. Above this temperature the PDF is characterized by a dominant
central peak whereas below this temperature there are two dominant
side peaks. In higher dimensions, this is characteristic of a first
order phase transition. Since the PDF contains all essential
thermodynamic information it is very important to be able to compute
it accurately: our approach provides a simple and accurate method for
computing thermodynamic quantities such as the specific heat, internal
energy, {\em etc.} A detailed description is now in preparation
\cite{HS}. 

\subsection{Field Configurations}

\begin{figure}
\centerline{\psfig{figure=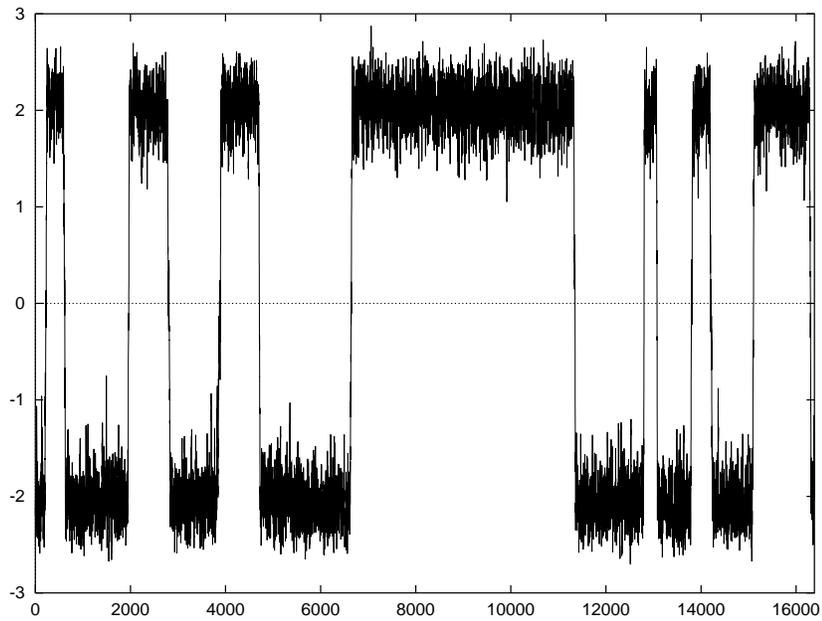,height=8cm,width=10cm}}
\caption[Figure 6]{\small{A sample field configuration at $1/\b=.2$
with all other parameters the same as in Fig. 5.}}
\end{figure}     

A low temperature field configuration is shown below in Fig. 6.  The
kinks are few in number but well defined. At higher temperatures the
number of kinks increases but their shape is smeared by thermal noise.
At still higher temperatures it becomes impossible to distinguish
these nonlinear structures from nonlinear phonons. At modest temperatures,
the shape of the kink computed from Langevin simulations can be compared
with the exact solutions described in Sec. 3 (for example, the kinks
in Fig. 6 correspond to the kinks of Fig. 1 with $\t\simeq .91$). Here
we simply report that the results from the simulations match well with
theoretical expectations. Details will be given elsewhere \cite{HS}.
To summarize, we have two different checks, namely the PDF and kink
shape, which both show very good accuracy and provide a high level of
confidence in our simulations. 

\subsection{Correlation Functions} 

The absolute value of the location of the side minima is the order
parameter for the first order phase transition. Interestingly, the
topological charge associated with a kink interpolating between these
two minima at the phase transition point ($T_c$, three degenerate
minima) turns out to be precisely equal to the order parameter
\cite{BK1}. In addition to the value of the order parameter, its
spatial correlations as well as correlations of its intensity are
often of interest in studying a phase transition \cite{KS}\cite{AHK}.
The correlation functions are particularly interesting because they
describe the behavior of systems which are nearly ordered but do not
undergo sharp phase transitions at any finite temperature. Following
the path integral (transfer operator) procedure \cite{KS}\cite{BK1},
the correlation functions are expressed in terms of the eigenvalues
and eigenstates of the transfer operator as follows:
\ba
C_1(x)&=&\sum_n\abs{\bra{\J_n}\F\ket{\J_0}}^2\exp\left[-\b{x\over a}
(\epsilon_n-\epsilon_0)\right]~,                         \label{c1}\\
C_2(x)&=&\sum_n\abs{\bra{\J_n}\d\abs{\F}^2\ket{\J_0}}^2
\exp\left[-\b{x\over a}(\epsilon_n-\epsilon_0)\right]~,  \label{c2}
\ea 
where $\b=1/k_BT$, $\d\abs{\F}^2=\abs{\F(x)}^2-\VEV{\abs{\F(x)}^2}$,
and $a$ is the lattice constant. 
 
For $x \gg \xi$, the lowest excited state coupled by the matrix
element determines the behavior of the correlation functions. In
other words, the eigenvalues set inverse correlation lengths. For $T
\simeq T_c$, near degeneracy in eigenstates is reflected in an
increased range of correlation (tunneling) \cite{KS}\cite{AHK}. At
large distances $C_1(x)$ and $C_2(x)$ are dominated by the state with
smallest eigenvalue for which the corresponding matrix elements are
nonvanishing (excluding the $n=0$ state). The correlation lengths for
$C_1$ and $C_2$ are, respectively,
\ba
{1\over \xi_1}&\simeq&{\b\over a}
(\epsilon_1-\epsilon_0)~,                               \label{xi1}\\  
{1\over \xi_2}&\simeq&{\b\over a}
(\epsilon_2-\epsilon_0)~.                               \label{xi2} 
\ea
$\xi_1$ is proportional to the average separation between neighboring
kinks, which is the distance over which the field remains correlated.
$\xi_1$ grows exponentially with decreasing temperature, and $\xi_1
\rightarrow \infty$ as $T \rightarrow 0$, when no kinks remain in the
system. The energy density correlations are usually short range. The
static structure factor (or the equal time correlation function) is
the Fourier transform of $C_1$, and is given by
\bq
S(q)={1\over 2\p}\int dx\hbox{e}^{iqx}\VEV{\F(0)\F(x)}. \label{str}
\eq
 
\begin{figure}
\centerline{\psfig{figure=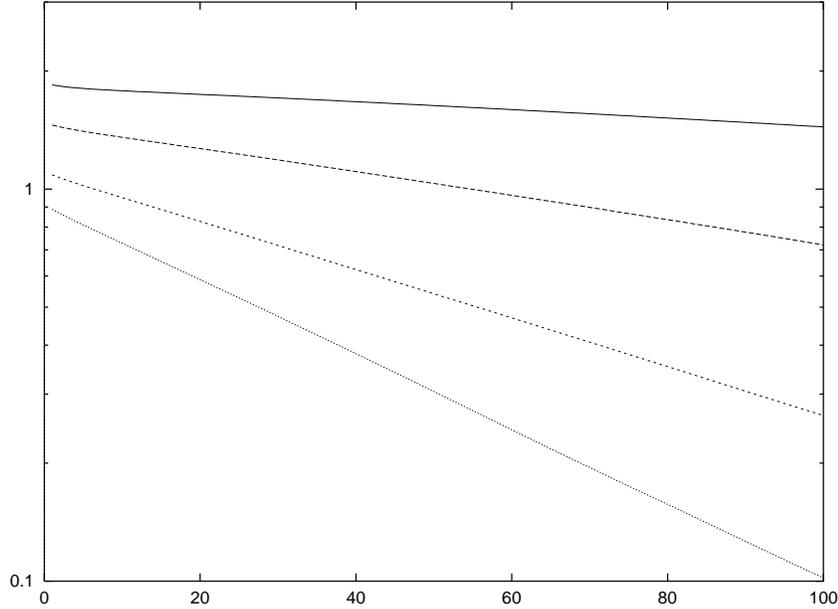,height=8cm,width=10cm}}
\caption[Figure 7]{\small{The (unnormalised) correlation function
$C_1$ at four different temperatures plotted on a logarithmic scale.
The parameters are the same as in Fig. 5, with only the temperature
being varied: from top to bottom $1/\b=.3,~.35,~.4,~.45$.}}
\end{figure}     

The field-field (or order-parameter-order-parameter) correlation
function $C_1$ for four temperatures is shown in Fig. 7. The
exponential decay is apparent. The correlation length 
is given directly by the slope of the correlation function plotted on
a logarithmic scale, while the average domain size in the system is
obtained from the first zero crossing.  The corresponding structure
factors (fast Fourier transform of $C_1$) are also easy to compute but
we do not display them here. More details on the correlation
functions, their exact and semi-exact calculation, and comparison with
numerical results will be given elsewhere \cite{HS}.

\section{Conclusion}
 
In conclusion we restate some key points. First, the $\F^6$ theory has
sufficient structure to describe first order phase transitions
especially relevant in the materials context (shape memory alloys)
\cite{FF}\cite{BK2}. A consequence of this complexity is the
appearance of several coherent nonlinear structures. The
thermodynamics of the theory can be profitably studied via both the
transfer operator method and Langevin simulations. The remarkable
occurrence of some exact solutions in the analytic transfer operator
approach for the $\F^6$ theory allows \cite{BK1} for a strong check on
the simulations. The very accurate determination of the PDF via our
simulations implies that this maybe a convenient window for a study of
the thermodynamics of such systems. Finally, extension of the Langevin
method to higher dimensions, and other classes of quasi-exactly
solvable potentials, is simple and we expect to present our results
for both the two and three dimensional cases soon.

\section{Acknowledgment}

We thank G. R. Barsch for fruitful discussions.
This work was supported by the U.S. Department of Energy at Los Alamos
National Laboratory. Numerical simulations were performed on the CM-5 at
the Advanced Computing Laboratory, Los Alamos National Laboratory.


\end{document}